# An alternative to common content management techniques

Rares Vasilescu

Computer Science and Engineering Department
Faculty of Automatic Control and Computers, Politehnica University
Bucharest, Romania

*Abstract*— **Content management systems use various strategies to store and manage information. One of the most usual methods encountered in commercial products is to make use of the file system to store the raw content information, while the associated metadata is kept synchronized in a relational database management system. This strategy has its advantages but we believe it also has significant limitations which should be addressed and eventually solved.**

**In this paper we propose an alternative method of storing and managing content aiming at finding solutions for current limitations both in terms of functional and nonfunctional requirements.**

*Keywords-CMS; content management; performance; architecture*

## I. INTRODUCTION

Content management systems (CMS) can be defined as a set of processes and technologies which support the digital information management lifecycle. This digital information is usually referred as "content" and can be found as not-structured or semi-structured - such as photographs, images, documents or XML data. While one can look at CMS as a software application, it is more and more used as a technological software platform on which other end-user applications are built. In turn, CMS are commonly based on other core technologies such as relational database management systems (RDBMS) and file systems thus is common for information and processes to traverse multiple technical layers to implement a given functionality.

The usage of out of the box components such as RDBMS helps systems achieve a lower time to market and high reliability metrics. On the other hand, this reuse comes with an inherent mismatch between components which can lead to nonoptimal performance, both in terms of functional and nonfunctional needs. Following experiments [3], [6] and practice we came to the conclusion that a high performance content management system needs to be designed specifically as an core infrastructure technology (such as database management systems are) rather than employing multiple layers from applications to data items.

We identified several key characteristics of CMS and during research and experiments each will be addressed and a new architecture implemented [8].

In Section 2 we will present such list of key functionalities, functionalities which should be addressed by a high performance implementation model. In Section 3 we will describe the proposed information storage alternative while in the next section we will discuss the challenges generated by this approach in terms of finding the managed data. The conclusion summarizes experimental results derived from the model implementation experience and from some performance benchmarks. It also outlines the next open points for research.

## II. CMS SPECIFIC FUNCTIONALITIES

During previous years, several efforts [1, 2] were made to standardize an interface to content management systems.

These initiatives have still some more room to expand but we can consider their existence as a validation of the fact that CMS becomes an infrastructure service, similar with database management systems and file systems. It therefore supports our approach of trying to design a high performance implementation model for CMS not necessarily based on other infrastructure services. In order to design a model for the CMS one must look at the key functions these systems provide and aim to implement them.

When looking at CMS functionalities set the following key features were identified:

- Data (content and metadata) management
- Security management
- Ability to ingest content
- Ability to process content
- Ability to classify content
- Retrieve data (metadata and content)
- Allow and control concurrent access
- Manage storage space
- Allow collaboration on content
- Allow definition of content enabled flows

We consider that each of these features can be explored from the point of view of high performance. The scope of this paper is not to address all of them but to present some first steps done in this direction and outlining the next activities which are done to build a high performance CMS. Understanding how content management systems are different from other systems (such as database management or file management systems) is essential for being able to design and build a high performance variant.

Content management usually needs a specialized approach on data management since it expresses a set of characteristics from which we mention the following:

- Manages complex data structures (not only data tuples with atomic values)

- Shows a high variance in each item data size

- Nonstructured content processing (e.g. text or image based search) is necessary for standard data discovery functions

- Security rules and management rules need to act at multiple levels on the complex data structures

### A. Complex data structures

Each element managed by such systems is comprised of a set of metadata (key-value(s) pairs) and the content itself (e.g. the binary file representing a photo).

Metadata are not only simple key-value pairs in which the value is an atomic element – they can also contain complex data structures sometimes repetitive (e.g. a table with many columns and rows). This characteristic is obviously in contradiction with the first normal form [5] and a relational database implementation will most probably not model it in this manner. But what we consider essential is that the actual information can be modeled in various ways and we should identify a method adequate for high performance.

Information also includes the actual data content which needs to be managed in synch with the metadata. There are ways for storing and managing this content inside the relational database tuples but experiments [3], [6] shown that such methods pose specific performance problems. Adding more challenge, each content item can have multiple versions which need to be tracked and managed. Versioning is not natively managed by common database management systems thus we can expect that such models are less than optimal. Content is not only versioned but can also be represented in multiple formats (each of the versions having multiple binary formats, such as multiple image representation formats of a picture). The relationship between renditions, versions and the information item itself should be addressed as core functionality.

### B. High variance of item size

Managed items vary substantially in size between CMS implementations and even inside the same system. It is not unusual to encounter a system with item size ranging from several bytes to multiple gigabytes or even terabytes.

A high performance system should address this characteristic at its core and provide means to efficiently store and manage each and every item with performance scaling at least linearly comparing with size.

### C. Nonstructured content processing

We are used to find and process information by using relational algebra on tuple based data organization. The fact that the piece of information is comprised of metadata and content at the same time leads to the need for at least enhancing the algebra with operators which can work on content. Since content is unstructured (or semi-structured in case or XML data, for example) such operators are different in nature than the common ones. Content processing is an essential function of CMS and is not unusual to be one of the most important functionality evaluated while choosing such a system. It is therefore mandatory that the system architecture embeds these at its core.

Another fact is that technology evolves while content not necessarily changes. For example a photo would be taken at a certain moment in time and its original representation remains the same while the manipulation technologies evolve and can extract and process more and more information based on the representation. Considering this, a CMS must allow this technological evolution without requiring a fundamental change and while still observing the performance topic.

### D. Security management

Arguably one of the top performance factors is the security model implementation subsystem. This is due to the fact that security should govern everything and this is not a trivial task to fulfill.

Each managed element usually has an associated security set which determines who can perform what kind of operation on it. Characteristic to CMS is that these security rules apply not only at item level but also at sub-item level. For example, one system user could have the permissions to update some of the document's metadata but not some of them and could operate on the content only for versioning not overwriting it. More, such permissions could address only an item version or format, not all of them (e.g. a user could be authorized to see only the PDF format of an item which also has a text editable format).

### III. PROPOSED STORAGE MODEL

The proposed model shows a content management system which stores data in an autonomous, self descriptive manner, scalable both in terms of functionality and of usage. Individual content items are self-described and stored in a standardized format on generic file systems. The file format (Fig. 1) can follow any container convention (e.g. can be XML based) but is essential to contain all the information necessary to manage that specific piece of content regardless of the software built for this reason.

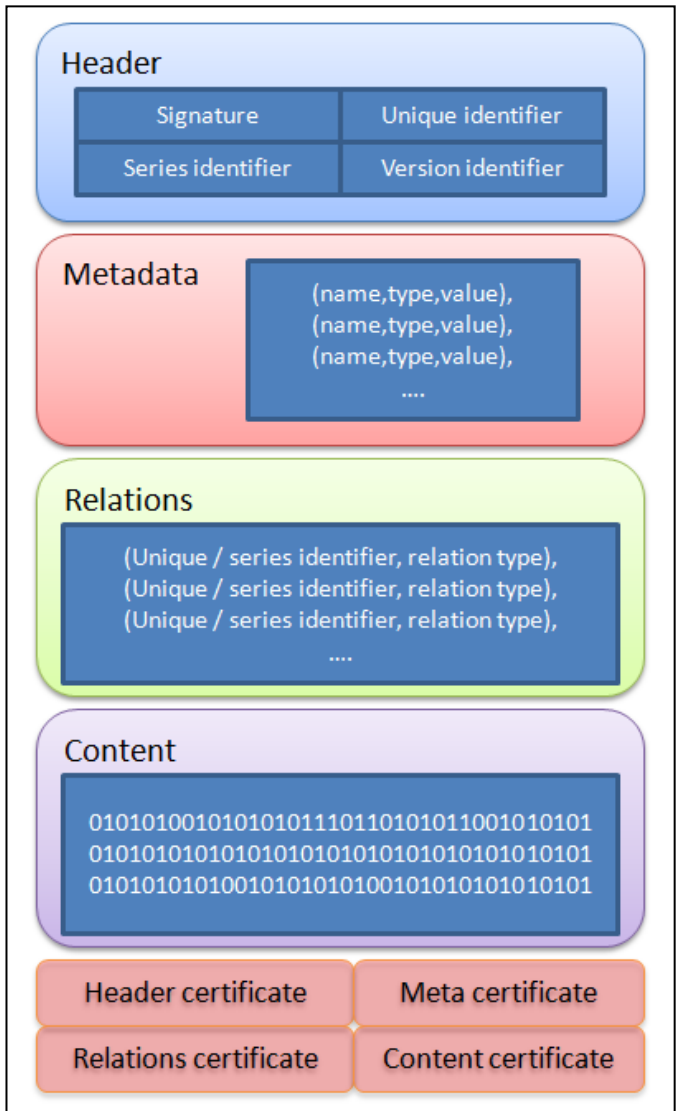

Figure 1.  Item file structure

The said file is designed to contain multiple segments, each representing a specific data area characterizing the information item. It is expected to store these segments in fixed size data pages (e.g. 1KB page increments) so that eventual updates do not trigger the rewrite of the entire file (which would be time consuming). Of course, the paging would increase the overhead on the storage space and this need to be considered when configuring the segment size. One option can be to define the segment size for each item or to dynamically choose it at system runtime based on item properties.

The header area begins the file and contains the format version and the key content item identifier. Alongside it must contain also the version series identifier and the version identifier. This makes each item very easy to identify without reading or processing the whole file. The strategy used to assign series identification is designed so it does not need an update of existing version metadata when a new version appears in the series, keeping existing items stable. It is essential to not need modifications into an existing item when related ones appear or are modified in turn. The main reason behind this requirement is that items can be stored also on read-only media (such as tapes or optical disks) and are therefore not-updateable physically. Also, compliance rules could mandate the read-only access and the requirement is thus not only from a technical limitation but also from a legal perspective.

Metadata values are contained into the next section (pairs of metadata name and associated values). A significant decision we took is to define each metadata completely (such as its name and data type) without referencing a global data-dictionary. This decision keeps the item self-described and independent of other data collections. The independence comes at the price of storage overhead since each metadata item which is present in several items is described also in each of them. This overhead would be significant if there would be a fixed schema by which to classify items. In exchange, we choose not to employ a schema based classification but to include in an item's metadata only the attributes which are relevant for that particular item. This decision has an impact also on the information retrieval techniques which need to be implemented since traditional methods are no longer suited.

Another section contains data about the links to other items. Each other item is referenced by unique identifier or by version series identifier. Each relation has also a type classification to differentiate between possible link variants. Relations are necessary to link together different versions of the same item and different formats of the same version.

After all these sections, the file records the content itself. This positioning of the data for several main reasons: any update of the metadata or associated relations can happen without accessing the whole file contents and the majority content updates can be handled by a file update not by an entire rewrite. In special cases we could choose to add at the end of the file some certificates to ensure the authenticity of item sections. These certificates can be applied using any kind of technique but one common method is using the IETF standard defined as RFC 3852 [7].

One addition to the above structure would be a header sub-section which can determine which other sections of the file are protected in a different manner than the others. For example, the actual content and a part of the metadata need to be read-only while some metadata information can be added or changed still. This is particularly useful for CMS compliance and retention needs.

## IV. SPECIFIC PROPOSED MODEL CHALLENGES

The proposed model is based on a series of architectural decision which have a significant impact on the overall system design. We will discuss here some impacted functionalities and propose ways of mitigating the risk of negative impact while enhancing the benefits.

Content is many times subject to retention rules. As the information gets transformed from physical supports (such as paper) to digital (e.g. scanned documents) the regulations also extend in similar ways. CMS users are expecting their system to provide methods of enforcing compliance rules and managing these processes as efficiently as possible. It is not uncommon for regulations to state that certain content be stored for many years (tens or even permanently) and on storage which prevent alterations (accidental or not). Looking back at current software application lifespan we can see that operating systems evolve significantly every several years and we can hardly see systems which remain stable over a decade. Given this state of things it is not reasonable to believe that a system built today will remain as is for the time needed to manage its content. With consideration to the above we proposed the item storage technique described in the previous section.

Each stored item has an essential ability of being self-described and stable during its lifecycle. Having items self-described empowers the system to employ information retrieval techniques which evolve in time while keeping the initial content unchanged. For example, when a new information processing technique is developed, the system can be simply extended to implement it also and then run it over the existing repository of items. More, the items can be stored on Write Once Read Many (WORM) mediums which can be stored outside the information system itself and processed only when needed (e.g. tapes libraries). All of this is possible by keeping the item catalog (index structure) separated to the content. The main difference versus common existing CMS models is that the catalog structure is not mandatory to be synchronized and maintained alongside the content itself since the content is self-described and such catalog can be entirely rebuilt only in a matter of time.

As previously presented, the self-described characteristic comes with an associated cost: overhead on the needed storage space and complexity of operations on the content file store itself generated by the paging algorithm. We believe that this cost is reduced since items do not include a fixed schema but are classified by individual characteristics (not even using a schema associated with item types). The approach gives the flexibility to classify an item by an initial set of attributes determined by the top application logic and then eventually add more metadata as the content progresses through its lifecycle. It helps a lot also in cases when an item needs to be perceived differently by various applications (e.g. a content item representing and invoice is used and classified differently by an accounts payable application then by a records management one). Considering that items have multiple versions and formats, this approach reduces significantly the metadata associated with each one since the only differentiating attributes can be stored on these items (e.g. format type) and the rest of them being inherited through the use of relations.

The current large majority of content management systems need to keep a data dictionary to describe each type of content item they manage. This might be seen as convenient for a number of system administration tasks but actually we found that it imposes a lot of restrictions and overhead. It is also naturally not flexible and a lot of workarounds need to be designed in order to allow concepts like "inheritance" or "aspects".

A challenge of the proposed model is to retrieve and make use of the stored items. Only storing the self-described items does not provide an efficient manner to access them by applying search filters – although this is possible with a full scan and filter approach. It is thus necessary to implement a data discovery mechanism which would enable application use the CMS for fast item retrieval and processing.

The proposed model considers also the lack of schema. Since there is no enforced schema, the top application is left with the task of choosing how an item is classified and then retrieved. Although this decision is different than the commonly established practice of the database system enforcing a schema which is then obeyed by caller applications we consider that this enforcement is necessary only when applications are not yet stable enough (e.g. in development phases) while afterwards the application itself becomes an enforcer of the schema. This assumptions is based on actual solution implementation experience and from observing that even though database systems have the ability to enforce referential constraints between tables, these features are seldom used when performance is key.

While it can be the task of the application to determine the metadata used for an item, it is still the task of the CMS to manage these data and to provide a way to filter it efficiently. We propose a system which includes a collection of independent agents, each of them processing an atomic part of data: a metadata field or a content piece. Once an item is created or updated, these agents get triggered and each of them processed and indexes the associated data. When search queries are submitted the filters will be splitted in basic operators and then submitted in parallel to respective search agents. These agents will process the sub-queries and return results as found to a query manager which aggregates the results and replies to the top application with partial results as they are collected.

A key advantage is that the top application can receive not only precise results but also results which partially match the search criteria. While this can seem not adequate (having a system which does not return precise matches) it can prove very efficient in practice since a user could be satisfied to obtain very fast an initial set of results and then – while it is evaluating the partial set – receive the complete result. One should note that the above terms "partial" and "complete" refer not only to the quantitative measure of the result (number of returned items) but also to the matching of partial or complete filter criteria.

A challenge to this model is the query optimization technique which cannot be based on traditional relational database models given the lack of schema and the related statistical information. Solving this challenge requires a change of the approach to optimization itself: not aiming to provide a complete response opens the door to other optimization techniques by focusing on the feedback from actual execution rather than preparing a query plan. This optimization should take into account the fact that given the vertical organization of the metadata (each agent having its own specific associated metadata item) the memory locality of frequently used index structures can help the process a lot. Since memory efficiency tends to grow at a faster pace than disk efficiency and processors tend to include multi-core elements more and more, we expect than an architecture geared up memory usage and parallel processing will provide the maximum possible performance right now.

Coming back to the item storage, a key element is the file organization within the file system itself. Tests [3] have shown that file systems generally provide very good performance but files need to be organized properly beforehand. For example, it is not advisable to store millions of file in the same "folder" inside a file system. While this is perfectly possible in modern file systems, it can experience major performance impact on accessing that folder – thus also any of the contained files. Although there are a lot of different file management systems available, this behavior is valid for most of them. The proposed solution is to store files in such way that the location of the file is determined by the unique object identifier and that no more than 256 files exist on the same folder. This is achieved by representing the unique identifier as a hexadecimal number resulting 8 pairs of 2 digits. The less significant pair represents the filename. The rest of the digits represent the folder names toward that content file (in order). By applying this simple logic files will not overwhelm file system folders and each item is directly identified on the disk, saving a lot of expensive I/O operations.

Other concerns of the proposed architecture are modern challenges such as refreshing digital signatures on fixed content for items which need long retention periods (e.g. over 10 years). For this reason, the content file has a dedicated area at the end of the file to store digital signatures on various areas (metadata and / or content). Multiple signatures can be stored for any area (e.g. successive signatures for same content part).

## V. Conclusion and next steps

Independent studies [4] show that about 80% of the stored data is not inside a database management system and that the total volume increases exponentially to reach over a thousand Exabytes by 2011 (ten times more than in 2006).

We believe that designing a CMS able to handle very large structured and semi structured content is key to maintaining the pace of this information growth. To validate the content storage techniques presented in high level within this paper, we work on building an actual implementation of the system and benchmarking it versus other common CMS products.

Since there is no known accepted benchmark procedure for content management systems we will consider the functional elements defined by industry standards such as CMIS [1] but we will also include nonfunctional requirements such as the ability of the system to manage information over extended time periods.

### AUTHORS PROFILE

Dipl. Eng. Rares Vasilescu is a PhD student at Politehnica University, Faculty of Automatic Control and Computers, Computer Science and Engineering Department, Bucharest, Romania.

Previous work includes studies and experiments on the performance of database management systems. Current research addresses the area of content management systems in preparation of the PhD thesis conclusion.